\documentclass{emulateapj}

\begin{document}

\shortauthors{Luhman et al.}
\shorttitle{Near-IR Detection of WD~0806-661~B}

\title{Near-infrared Detection of WD~0806-661~B
with the {\it Hubble Space Telescope}\altaffilmark{1}}

\author{
K. L. Luhman\altaffilmark{2,3},
C. V. Morley\altaffilmark{4},
A. J. Burgasser\altaffilmark{5},
T. L. Esplin\altaffilmark{2},
and J. J. Bochanski\altaffilmark{6}
}

\altaffiltext{1}{Based on observations made with the NASA/ESA
{\it Hubble Space Telescope}
through program 12815, obtained at the Space Telescope Science Institute,
which is operated by the Association of Universities for Research in
Astronomy, Inc., under NASA contract NAS 5-26555, and 
observations with the ESO Telescopes at Paranal Observatory
under programs ID 089.C-0428 and ID 089.C-0597.
}
\altaffiltext{2}{Department of Astronomy and Astrophysics, The Pennsylvania
State University, University Park, PA 16802, USA; kluhman@astro.psu.edu}
\altaffiltext{3}{Center for Exoplanets and Habitable Worlds, The 
Pennsylvania State University, University Park, PA 16802, USA}
\altaffiltext{4}{Department of Astronomy and Astrophysics, University of
California, Santa Cruz, CA 95064, USA}
\altaffiltext{5}{Center for Astrophysics and Space Science, University of
California San Diego, La Jolla, CA 92093, USA}
\altaffiltext{6}{Haverford College, 370 Lancaster Avenue, Haverford, PA 19041,
USA}

\begin{abstract}

WD~0806-661~B is one of the coldest known brown dwarfs
($T_{\rm eff}=300$--345~K) based on previous mid-infrared photometry
from the {\it Spitzer Space Telescope}. In addition, it is a
benchmark for testing theoretical models of brown dwarfs because 
its age and distance are well-constrained via its primary star
(2$\pm$0.5~Gyr, 19.2$\pm$0.6~pc).
We present the first near-infrared detection of this object, which has
been achieved through F110W imaging ($\sim Y+J$) with the Wide Field
Camera 3 on board the {\it Hubble Space Telescope}.
We measure a Vega magnitude of $m_{110}=25.70\pm0.08$, which implies
$J\sim25.0$.
When combined with the {\it Spitzer} photometry, our estimate of $J$
helps to better define the empirical sequence of the coldest brown dwarfs
in $M_{4.5}$ versus $J-[4.5]$.
The positions of WD~0806-661~B and other Y dwarfs in that diagram are
best matched by the cloudy models of \citet{bur03} and the cloudless models
of \citet{sau12}, both of which employ chemical equilibrium.
The calculations by \citet{mor14} for 50\% cloud coverage differ only modestly
from the data.
Spectroscopy would enable a more stringent test of the models, but based
on our F110W measurement, such observations are currently possible only with
{\it Hubble}, and would require at least $\sim10$ orbits to reach a
signal-to-noise ratio of $\sim5$.

\end{abstract}

\keywords{
binaries: visual --- 
brown dwarfs ---
infrared: planetary systems --- 
infrared: stars ---
planets and satellites: atmospheres}

\section{Introduction}
\label{sec:intro}

Wide-field near-infrared (IR) imaging surveys have found hundreds of brown
dwarfs in the solar neighborhood, reaching objects near the end of the
T spectral sequence
\citep[$T_{\rm eff}>500$~K,][]{bur04,war07,del08,burn10,luc10,alb11}.
Two members of the colder Y spectral class \citep{cus11} 
have been discovered as close companions to T dwarfs through
near-IR adaptive optics imaging \citep{liu11,liu12}.
However, because the near-IR fluxes of brown dwarfs collapse below
$\sim400$~K \citep{bur03}, space-based mid-IR telescopes
offer the best sensitivity to Y dwarfs.
Fortunately, two telescopes of this kind have been available in recent years,
consisting of the {\it Spitzer Space Telescope} \citep{wer04}
and the {\it Wide-field Infrared Survey Explorer} \citep[{\it WISE},][]{wri10}.
{\it Spitzer} imaging of fields surrounding nearby stars has uncovered a
companion that is likely a Y dwarf \citep{luh11} and the
all-sky map from {\it WISE} has found 19 free-floating objects that
are confirmed or likely Y dwarfs \citep{cus11,cus14,kir12,kir13,tin12,luh14,pin14}.
Following the initial mid-IR detections of Y dwarfs by these satellites,
further characterization of their spectral energy distributions is currently
possible only through deep photometry or spectroscopy at near-IR wavelengths
\citep{leg13,cus14}.

The probable Y dwarf discovered with {\it Spitzer} is a companion
to the white dwarf WD~0806-661. It has an estimated mass of
6--9~$M_{\rm Jup}$ \citep{luh12} and a projected separation of $130\arcsec$,
corresponding to 2500~AU at the distance of the primary 
\citep[19.2$\pm$0.6~pc,][]{sub09}.
WD~0806-661~B has not been classified spectroscopically, but
it is likely to have a Y spectral type since it is less luminous
than most confirmed Y dwarfs \citep{luh12,mar13,dup13,bei14}. 
In addition to being one of the coldest known brown dwarfs
\citep[$T_{\rm eff}=300$--345~K,][]{luh12}, WD~0806-661~B
was the first Y dwarf with an accurate distance because of
its companionship to a star with a previously measured parallax, and it
is the only Y dwarf with a well-constrained age because its primary is a white
dwarf. As a result, it represents a unique benchmark for testing atmospheric
and evolutionary models of substellar objects at very low temperatures.
However, because of its large distance
compared to other known Y dwarfs, WD~0806-661~B is particularly faint
and few data are available for comparison to models;
it has been detected in only the 3.6 and 4.5~\micron\ bands of {\it Spitzer}
to date and has a rather faint limit on its near-IR flux
\citep[$J>23.9$,][]{luh12}.

In pursuit of the first near-IR detection of WD~0806-661~B, we have 
obtained images of it with the Wide Field Camera 3 (WFC3) on board the
{\it Hubble Space Telescope} ({\it HST}), which is the most sensitive near-IR
camera that is currently available. We have also analyzed archival images of
WD~0806-661~B from the Very Large Telescope (VLT). In this paper, we describe
the collection and reduction of these data (Section~\ref{sec:obs})
and use the reduced images to constrain the binarity of WD~0806-661~B,
refine the empirical sequence of Y dwarfs in a color-magnitude diagram,
and test models of brown dwarfs (Section~\ref{sec:analysis}).
We conclude by placing our near-IR data for WD~0806-661~B in the context
of other known Y dwarfs and discussing additional observations 
that would further constrain the models (Section~\ref{sec:disc}).

\section{Observations}
\label{sec:obs}

\subsection{F110W Images from {\it HST}}

We obtained images of WD~0806-661~B with the IR channel of WFC3 \citep{kim08}
on 2013 February 8 and 9 (UT).
The camera contains a $1024\times1024$ HgCdTe array in which the pixels
have dimensions of $\sim0\farcs135\times0\farcs121$.
The inner $1014\times1014$ portion of the array detects light, which
corresponds to a field of view of $136\arcsec\times123\arcsec$.
We selected the F110W filter for these observations because it appeared to
offer the best sensitivity to cold brown dwarfs among the available
WFC3 filters based on simulations with the instrument's exposure time
calculator using spectra of late T dwarfs.
This filter spans from $\sim0.9$--1.4~\micron, corresponding roughly to
the sum of the $Y$ and $J$ filters.
The observations were performed during three identical two-orbit visits.
In a given orbit, one 1003~s exposure was collected at each position in
a three-point dither pattern. The dither patterns in the two orbits in
each visit were offset by 3.5 pixels along the x-axis of the array.

\begin{figure}
\epsscale{0.5}
\plotone{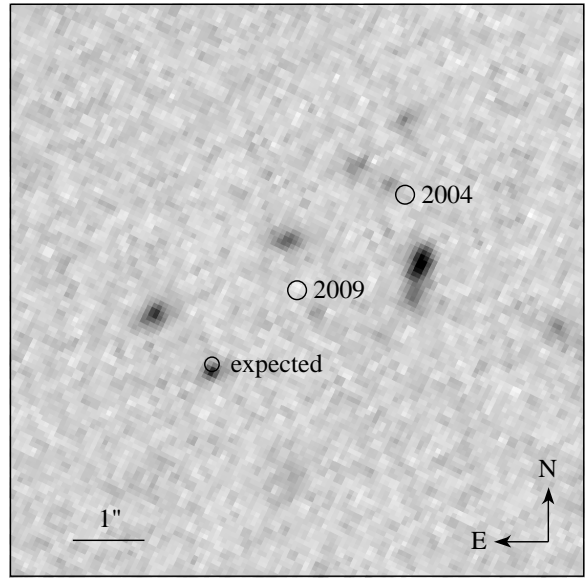}
\caption{
WFC3 F110W image of a $8\arcsec\times8\arcsec$ field encompassing
WD~0806-661~B. We have marked the positions of this object measured with
{\it Spitzer} in 2004 and 2009 \citep{luh11} and the position
expected in this image based on those earlier detections and the
proper motion and parallax of the primary. A source is detected
near the expected location, which we take to be WD~0806-661~B.
The radius of each circle corresponds to 1~$\sigma$. 
}
\label{fig:image}
\end{figure}

The 18 WFC3 images were registered and combined using the tasks
{\it tweakreg} and {\it astrodrizzle} within the DrizzlePac software package.
We adopted a drop size of 0.85 native pixels
and a resampled plate scale of $0\farcs065$~pixel$^{-1}$.
Point sources in the reduced image exhibit FWHM$\sim0\farcs18$.
We aligned the world coordinate system of the WFC3 image to that of
the IRAC images from \citet{luh11,luh12} using offsets in
right ascension, declination, and rotation that were derived from sources
detected by both WFC3 and IRAC. In Figure~\ref{fig:image}, we
present a small portion of the WFC3 image surrounding the positions
of WD~0806-661~B in the IRAC images, which were taken in 2004 and 2009.
We have estimated the location of WD~0806-661~B on the date
of the WFC3 observation by combining the astrometry from IRAC with the
proper motion and parallax of the primary \citep{sub09}.
As shown in Figure~\ref{fig:image}, a source is detected in the WFC3 image at
$\sim1$~$\sigma$ from the expected position, which we take to be WD~0806-661~B.

Aperture photometry was measured for WD~0806-661~B using 
an aperture radius of $0\farcs26$ and radii of $0\farcs26$ and $0\farcs65$
for the inner and outer boundaries of the sky annulus, respectively.
We converted that measurement to an aperture radius of $0\farcs4$ using
an aperture correction of 0.097~mag, which was estimated from well-detected,
isolated stars in the image.
We then applied the zero-point Vega magnitude for F110W for an aperture radius
of $0\farcs4$, which is
25.8829\footnote{http://www.stsci.edu/hst/wfc3/phot\_zp\_lbn}.
The resulting F110W photometry is given in Table~\ref{tab:data}.

\subsection{$z$ and $Y$ Images from VLT}

To further constrain the spectral energy distribution
of WD~0806-661~B, we have made use of unpublished images in the $z$ and
$Y$ bands that are publicly available in the data archive of the VLT.
The $z$-band images were collected with the Focal Reducer/Low
Dispersion Spectrograph 2 \citep{app98} on the VLT Unit Telescope 1
on the night of 2012 March 30 (P. Delorme, ID 089.C-0597).
The camera contains two 2048$\times$4096 CCDs.
The observations were performed with the Standard Resolution collimator and
2$\times$2 binning, which produced a plate scale of $0\farcs25$ pixel$^{-1}$.
WD~0606-661~B was placed within one array and 38 dithered images were taken,
each with an exposure time of 120~s.
We bias subtracted and flat fielded the individual frames,
and registered and combined them into a single mosaic.
The FWHM of point sources in the combined image was $\sim0\farcs7$.
The image was flux calibrated using photometry in $z$(AB) for a calibration
star, G138-31, from the Ninth Data Release of the Sloan Digital Sky Survey
Photometric Catalog \citep{ahn12} combined with a conversion from $z$(AB) to
$z$(Vega)\footnote{http://www.sdss.org/dr7/algorithms/sdssUBVRITransform.html}.

The $Y$-band images were obtained with the High Acuity Wide-Field K-band Imager
\citep{kis08} on the VLT Unit Telescope 4 on the night of 2012 May 21
(M. Burleigh, ID 089.C-0428).
The camera contains four 2048$\times$2048 HgCdTe arrays and has a plate scale
of $0\farcs106$ pixel$^{-1}$.
WD~0806-661~B was placed within one of the four arrays and 28 dithered
images were taken, each consisting of 26 coadded 2~s exposures.
Following dark subtraction and flat fielding, the images were registered
and combined. Point sources in the combined image had FWHM$\sim0\farcs6$.
Stars in the observed field exhibited significant flux variations
($\sim$0.4~mag) among the dithered frames, indicating that the conditions
were not photometric and that the photometric standard star probably
would not provide an accurate flux calibration.
For stars appearing in our $z$ and F110W images and the previous $J$-band
images from \citet{luh12}, the median values of $z-m_{110}$ and $m_{110}-J$
differ by $\sim0.2$~mag; the difference in the medians of
$Y-m_{110}$ and $m_{110}-J$ should be even smaller because of the smaller
wavelength range spanned.
Therefore, we performed a rough calibration of the $Y$-band image by
requiring that $Y-m_{110}$ and $m_{110}-J$ have the same medians.

We find that WD~0806-661~B is not detected in the $z$ and $Y$ images.
The magnitude limits that correspond to a signal-to-noise ratio (S/N) of
3 are provided in Table~\ref{tab:data}.

\section{Analysis}
\label{sec:analysis}

\subsection{Multiplicity Constraints}

\begin{figure}
\epsscale{0.6}
\plotone{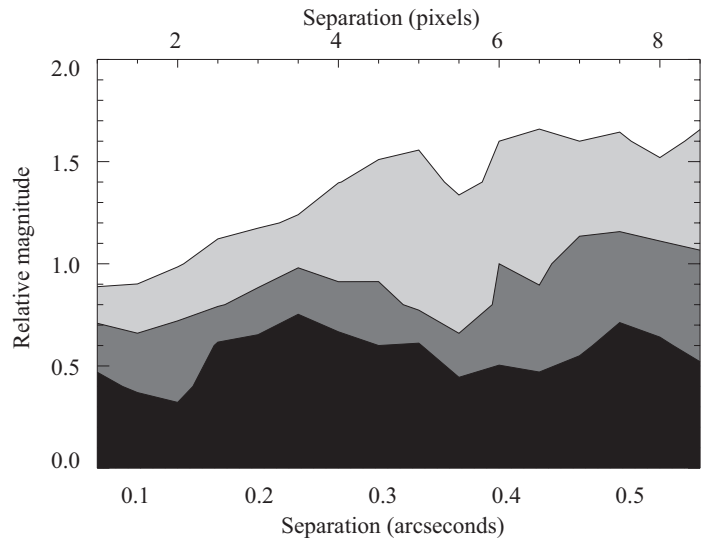}
\caption{
Relative magnitude limits at which 20\%, 50\%, and 80\% (top to bottom)
of simulated companions to WD~0806-661~B are recovered by our PSF analysis
of the WFC3 F110W image.
}
\label{fig:bin}
\end{figure}

We searched for binary companions to WD~0806-661~B in the drizzle-combined
F110W image through a point source function (PSF) fitting algorithm.
An empirical PSF was constructed by median-combining normalized and
background-subtracted $1\farcs4\times1\farcs4$ subimages of 96 high
S/N ($>$50), unsaturated, normalized point sources in the drizzled
frame. This PSF may not precisely represent those of WD~0806-661~B
or a cooler companion due to the significantly different spectral energy
distributions of the field stars; however, this is our best option given the
necessity of using the drizzle-combined image. An alternate method,
using Tiny Tim PSF models \citep{kri95}, is also not feasible
as this code is currently not set up for subsampled WFC3 images. We fit both
single and binary PSF models to a comparably-sized subimage centered on
WD~0806-661~B using a Monte Carlo Markov Chain algorithm with $\chi^2$
evaluation to determine the optimal primary and secondary $x$ and $y$
pixel positions and fluxes.  The binary model failed to provide a significantly
better fit ($>$90\% confidence) to the data based on an F-test analysis,
so we rule out detection of a companion. We assessed our sensitivity limit
by performing an identical series of binary fits to 10,000 simulated subimages
with companion PSFs implanted, sampling $0<\Delta m_{110}<3$, separations
of 1 to 8.5 pixels ($0\farcs065$--$0\farcs553$) and all position angles.
Fits were deemed successful if the
recovered companion had a separation and relative magnitude within 0.5~pixels
and 0.3~mag of the input parameters.  Figure~\ref{fig:bin} displays the 20\%,
50\% and 80\% recovery limits as a function of separation and relative
magnitude ($\Delta m_{110}$). Beyond $0\farcs13$, the 50\% (80\%) limit is 
$\Delta m_{110}\sim0.9$~mag ($\sim0.7$~mag), which corresponds to
$\Delta$T$_{eff}\sim15$~K based on the models of brown dwarfs
described in Section~\ref{sec:models}.
These limits were verified by visual examination of the implanted companions.

\subsection{Comparison of Observed and Model Photometry}
\label{sec:compare}

\subsubsection{Y Dwarf Models Selected for Comparison}
\label{sec:models}

For comparison to the data for WD~0806-661~B, we have considered four
sets of models for the spectral energy distributions of Y dwarfs that are
characterized primarily by the following features:
water clouds and chemical equilibrium \citep{bur03}, 
cloudless and chemical equilibrium \citep{sau08,sau12}, 
cloudless and non-equilibrium chemistry \citep{sau08,sau12}, and
50\% coverage of water, chloride, and sulfide clouds and chemical equilibrium
\citep{mor12,mor14}.
These models have been previously compared to color-magnitude diagrams
of late T and Y dwarfs, the results of which can be summarized as follows:
$Y-J$ is better matched by the cloudless models while the cloudy models predict
colors that are too red \citep{liu12,leg13,mor14}.
The opposite is true for $J-H$, which is reproduced by the cloudy models
of \citet{mor12,mor14} while the cloudless model colors are too blue
\citep{mor12,mor14,leg13,mar13,bei14}.
The values of $H-K$ predicted by both cloudy and cloudless models are
too blue, possibly because of incomplete methane line lists \citep{mor14}.
The mid-IR bands that have been considered consist of the 3.4, 4.6, and
12~\micron\ bands from {\it WISE} ($W1$, $W2$, and $W3$) and the 3.6 and
4.5~\micron\ bands from {\it Spitzer} ([3.6] and [4.5]).
Both the cloudy and cloudless models are roughly consistent with previous
data for Y dwarfs in color-magnitude diagrams composed of [4.5] or $W2$
and colors spanning from those bands to $J$, $H$, and $W3$ \citep{leg13,mor14},
although the large scatter in the data precludes discrimination between the
models. Finally, both sets of model colors are too red in color-magnitude
diagrams involving $[3.6]-[4.5]$ \citep{leg10a,luh12,bei14}.

\subsubsection{Color-Magnitude Diagrams}
\label{sec:cmd}

The previous photometry and our new measurements for WD~0806-661~B
are compiled in Table~\ref{tab:data}.
We wish to use our new data to test the models of Y dwarfs described in
the previous section. Before doing so, we would like to convert $m_{110}$
to $J$ since it is more widely available for other Y dwarfs and is encompassed
by the F110W filter. To perform this conversion, we have computed
$m_{110}-J$ from observed and model spectra of cold brown dwarfs using
the transmission profiles of F110W and $J$ on the Mauna Kea Observatories
Near-Infrared system \citep{sim02,tok02,tok05}.
For the observed spectra, we have considered the published spectra of
dwarfs later than T8 that fully span the wavelength range of F110W and that
have the highest S/N, which corresponds to the data for
UGPS J072227.51-054031.2 (T9), WISEPC J014807.25-720258.8 (T9.5),
and WISEP J154151.65-225025.2 (Y0.5) from \citet{cus11}.
These spectra exhibit $m_{110}-J=0.88$, 0.87, and 0.65, respectively.
Meanwhile, the models that are near the value of $M_{4.5}$ for
WD~0806-661~B (similar to that of WISEP J154151.65-225025.2) produce
$m_{110}-J$ colors of 1.2 \citep{bur03}, 0.66 (chemical equilibrium) and
0.81 (non-equilibrium) \citep{sau12}, and 0.75 \citep{mor14}.
We ignore the color from \citet{bur03} since it is significantly redder than
the data for T/Y dwarfs 
\citep[a similar difference is present in $Y-J$,][]{liu12}.
The remaining model colors are roughly similar to the
observed values. These observed and model colors suggest that WD~0806-661~B
probably has $m_{110}-J\sim0.7$, which implies $J\sim25.0$ when
combined with our measurement of $m_{110}$.

\begin{figure}
\epsscale{1}
\plotone{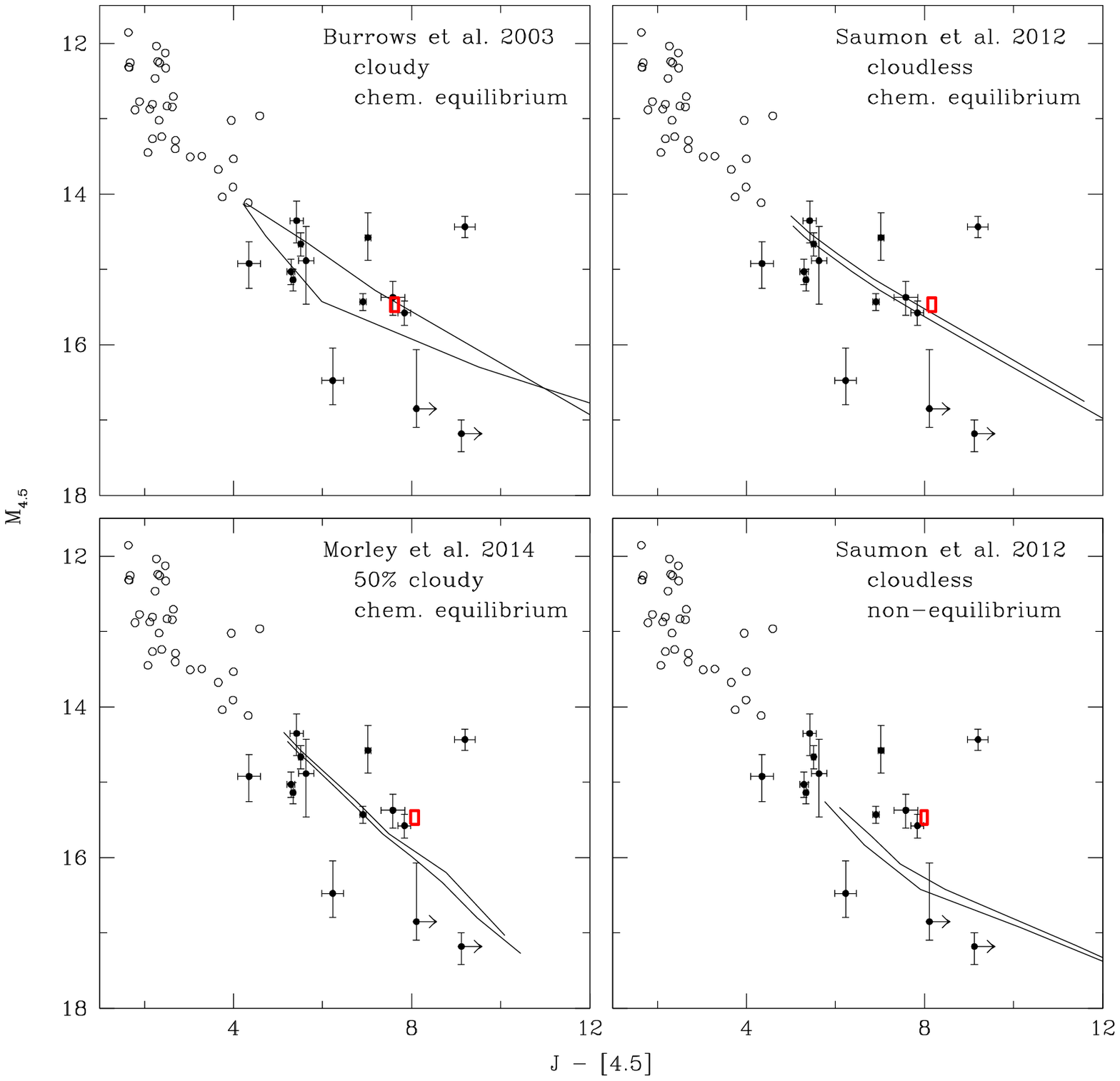}
\caption{
Color-magnitude diagrams for WD~0806-661~B (open rectangle)
and samples of T dwarfs \citep[open circles,][references therein]{dup12}
and Y dwarfs \citep[filled circles with error
bars,][]{cus11,cus14,tin12,bei13,bei14,leg13,mar13,kir13,dup13,luh14}. 
These data are compared to the magnitudes and colors predicted
by four sets of models for ages of 1 and 3~Gyr (solid lines), which
encompass the age of WD~0806-661~B \citep[$2\pm0.5$~Gy,][]{luh12}.
}
\label{fig:p1}
\end{figure}

\begin{figure}
\epsscale{1}
\plotone{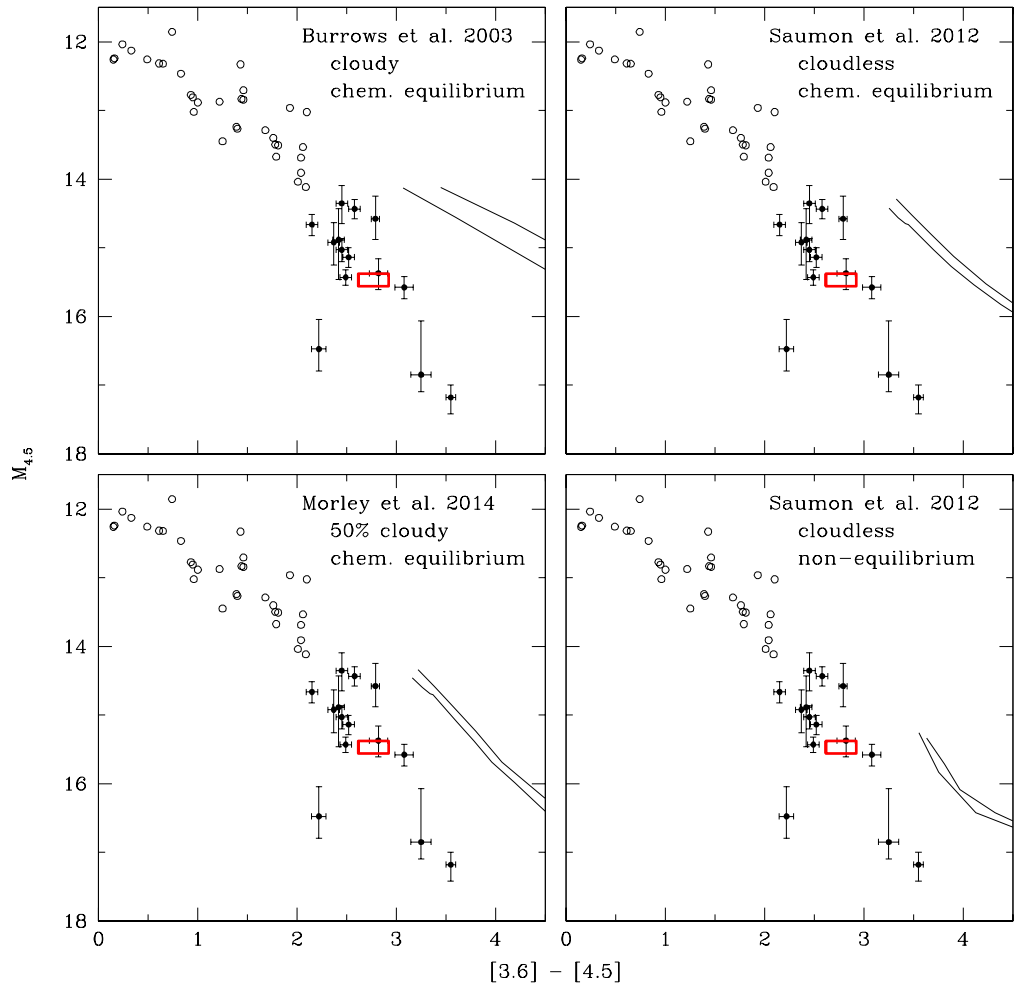}
\caption{
Same as Figure~\ref{fig:p1}, except that the color is $[3.6]-[4.5]$.
}
\label{fig:p2}
\end{figure}

We can derive new constraints on the colors of WD~0806-661~B by combining
our estimate of $J$, the limits on $z$ and $Y$, and the existing
photometry in [4.5]. These data produce $z-J>-0.4$, $Y-J>-1.8$, and
$J-[4.5]\sim8.1$. For the first two colors, most previous measurements
for Y dwarfs range from $z-J=2.2$ to 3.2 (with Vega $z$) and from
$Y-J=-0.5$ to 0.4 \citep{liu12,leg13,lod13}. Thus, the $z$ and $Y$ data are not
sufficiently deep to provide useful constraints on those colors.
To make use of $J-[4.5]$, we plot WD~0806-661~B on a diagram
of $M_{4.5}$ versus $J-[4.5]$ in Figure~\ref{fig:p1}.
We choose $M_{4.5}$ as the magnitude since this band encompasses less
atmospheric absorption and usually has the smallest photometric errors among
the broad-band filters in which Y dwarfs have been observed.
WD~0806-661~B is also shown on a diagram of $M_{4.5}$ versus $[3.6]-[4.5]$
in Figure~\ref{fig:p2}. In both diagrams, we include a sample of T dwarfs
\citep{dup12} and all other known Y dwarfs that have photometry in these
bands and parallax measurements
\citep{cus11,cus14,tin12,bei13,bei14,leg13,mar13,kir13,dup13,luh14}.
The data are shown separately with each of the four sets of Y dwarf models
that we are considering (Section~\ref{sec:models}) for ages of 1 and 3~Gyr,
which bracket the age of WD~0806-661 \citep[2$\pm$0.5~Gyr,][]{luh12}.
In the diagram of $M_{4.5}$ versus $J-[4.5]$ for a given set of models,
$J$ for WD~0806-661~B has been estimated by combining our measurement
of $m_{110}$ with the value of $m_{110}-J$ predicted by those models
at the 4.5~\micron\ absolute magnitude of WD~0806-661~B.
Thus, the relative positions of WD~0806-661~B and the models in that
diagram are as they would be in a diagram of $M_{4.5}$ versus $m_{110}-[4.5]$.

The Y dwarfs in $M_{4.5}$ versus $J-[4.5]$ in Figure~\ref{fig:p1}
exhibit significant scatter, but the data have reached sufficient quality
that a recognizable sequence is becoming apparent among a subset of Y dwarfs.
This sequence consists of the five Y dwarfs at $J-[4.5]\sim5$--6,
the four Y dwarfs at $J-[4.5]\sim7$--8, and WD~0806-661~B.
The vertical dispersion among these objects is similar to that of the T dwarf
sequence. The scatter in $M_{4.5}$ versus $[3.6]-[4.5]$ is smaller 
and the Y dwarf sequence is better defined, as shown in Figure~\ref{fig:p2}.
WISE J014656.66+423410.0 is a moderate outlier in $M_{4.5}$ versus $J-[4.5]$
(the bluest Y dwarf), but it is within the sequence of other Y dwarfs
in $M_{4.5}$ versus $[3.6]-[4.5]$. WISE J035934.06$-$540154.6 falls
below the Y dwarf sequence in both diagrams.
WISE J071322.55-291751.9 and WISE J140518.40+553421.5 appear to be
overluminous in each diagram to a degree that is consistent with
unresolved binaries. WISE J182831.08+265037.8 is also brighter than
both sequences, although the difference in $M_{4.5}$ versus $J-[4.5]$ is
too large to be explained through binarity alone. The anomalous photometric
properties of this object have been discussed previously \citep{bei13,leg13}.
WD~0806-661~B does not appear to be overluminous in either color-magnitude
diagram, indicating that it is probably not an unresolved binary in which the
components have similar fluxes.

In Figures~\ref{fig:p1} and \ref{fig:p2}, the Y dwarf isochrones at 1 and
3~Gyr from \citet{bur03} differ only modestly; the isochrones for these ages
are quite similar to each other for each of the remaining three sets of models.
Because the model isochrones do not vary significantly with age, we
can compare them to both WD~0806-661~B and the Y dwarf population as a whole.
In $M_{4.5}$ versus $J-[4.5]$, the cloudy models of \citet{bur03} and
the cloudless models of \citet{sau12} agree fairly well with 
WD~0806-661~B and the Y dwarf sequence. The cloudy models of \citet{mor14}
are slightly bluer than those data at fainter magnitudes, and the
non-equilibrium models of \citet{sau12} are bluer still.
In $M_{4.5}$ versus $[3.6]-[4.5]$, all of the models are significantly redder
than the data, which has been noticed previously for T and Y dwarfs
\citep{leg10a,bei14}.
Given that the models resemble the data in $M_{4.5}$ versus $J-[4.5]$
but not $M_{4.5}$ versus $[3.6]-[4.5]$, this discrepancy may be
primarily due to errors in the predicted 3.6~\micron\ fluxes, although
near- and mid-IR spectroscopy are needed for a definitive conclusion.

\section{Discussion}
\label{sec:disc}

Through imaging with WFC3 on {\it HST}, we have obtained the first 
near-IR detection of the coldest known benchmark brown dwarf, WD~0806-661~B.
We have measured $m_{110}=25.7$, from which we estimate $J\sim25.0$.
When combined with previous {\it Spitzer} photometry, these data
produce $J-[4.5]\sim8.1$, which makes WD~0806-661~B approximately
the fourth reddest known Y dwarf in that color.
WD~0806-661~B also could be compared to other Y dwarfs in terms of $M_J$ to
constrain their relative temperatures, although that is better done
with the more accurate photometry that has been measured at 4.5~\micron\ for
most Y dwarfs. However, mid-IR photometry is unavailable for two Y dwarfs
because they are close companions, consisting of CFBDSIR J1458+1013~B and
WISEPC J121756.91+162640.2~B \citep{liu11,liu12}.
WD~0806-661~B is $\sim30$ times fainter than those objects in $M_J$, which
indicates that it is $\sim100$~K colder based on models of Y dwarfs.
In $J-[4.5]$ versus $M_{4.5}$, WD~0806-661~B has a relatively well-constrained
position that helps to solidify an empirical sequence of Y dwarfs that
has recently begun to emerge.

We have compared the sequence produced by WD~0806-661~B and other Y dwarfs
in $J-[4.5]$ versus $M_{4.5}$ to the predictions of theoretical models.
The cloudy models of \citet{bur03} and the cloudless models of \citet{sau12}
provide the best agreement with the data, although only modest differences
are present with the cloudy models of \citet{mor14}. Based on these
results and previous comparisons in other color-magnitude
diagrams \citep[e.g.,][]{leg13,mor14}, no single set of models
produces a clearly superior fit to the spectral energy distributions
of Y dwarfs at this time.
According to \citet{sau12} and \citet{mor14}, the differences in the
$J-[4.5]$ colors for cloudy and cloudless atmospheres increase with
fainter magnitudes. Thus, measurements of $J-[4.5]$ for the few known Y dwarfs
that are less luminous than WD~0806-661~B would help to discriminate between
those models. For instance, WISE J085510.83$-$071442.5 is the coldest known
brown dwarf ($T_{\rm eff}\sim250$~K) and has not been detected at near-IR
wavelengths \citep[$J>23$,][]{luh14}. The cloudy and cloudless models
of \citet{sau12} and \citet{mor14} predict that it should have $J-4.5\sim10.3$
and 12.7, corresponding to $J\sim24.2$ and 26.6, respectively.
Other commonly considered colors like $Y-J$ and $J-H$ do not differ
significantly between those models for the coldest Y dwarfs
\citep{mor14}, but measurements of these colors for WD~0806-661~B
and colder objects nevertheless would be useful for further constraining
the models in general.

Spectroscopy of WD~0806-661~B will be necessary in order to
fully exploit its potential as a benchmark for testing model atmospheres.
Our near-IR photometry indicates that it is too faint for spectroscopy
with existing ground-based telescopes. WFC3 on {\it HST} is the only available
spectrograph that
is capable of detecting it. Based on previous WFC3 observations of Y dwarfs
\citep{cus11,cus14,kir12,kir13}, a minimum of $\sim10$ orbits would be
required to reach S/N$\sim5$ in low-resolution near-IR spectroscopy of
WD~0806-661~B. 
Because they will offer greater near-IR sensitivity than {\it HST} and will
extend to mid-IR wavelengths, the next generation of 20--40~m ground-based
telescopes and the {\it James Webb Space Telescope} will enable detailed
characterization of the spectra of WD~0806-661~B and other Y dwarfs and
discrimination among the competing models of their atmospheres \citep{mor14}.

\acknowledgements
We acknowledge support from grant GO-12815 from the Space Telescope Science
Institute. The Center for Exoplanets and Habitable
Worlds is supported by the Pennsylvania State University, the Eberly College
of Science, and the Pennsylvania Space Grant Consortium.

\begin{deluxetable}{lll}
\tabletypesize{\scriptsize}
\tablewidth{0pt}
\tablecaption{Photometry of WD~0806-661~B\label{tab:data}}
\tablehead{
\colhead{Band} & \colhead{Magnitude} & \colhead{Reference}
}
\startdata
$z$ & $>$24.6 & 1 \\
$Y$ & $>$23.2 & 1 \\
$m_{110}$ & 25.70$\pm$0.08 & 1 \\
$J$ & $>$23.9 & 2 \\
$J3$ & $>$23.5 & 2 \\
$[3.6]$ & 19.65$\pm$0.15 & 2 \\
$[4.5]$ (2004) & 16.96$\pm$0.09 & 2 \\
$[4.5]$ (2009) & 16.84$\pm$0.06 & 2 \\
$[4.5]$ (mean) & 16.88$\pm$0.05 & 2 \\
\enddata
\tablecomments{All data are Vega magnitudes. The limits correspond to S/N$>$3.
Observed and model spectra of Y dwarfs imply $m_{110}-J\sim0.7$
(Section~\ref{sec:cmd}).}
\tablerefs{
(1) this work;
(2) \citet{luh12}.}
\end{deluxetable}

\clearpage

\begin{thebibliography}{}

\bibitem[Ahn et al.(2012)]{ahn12}
Ahn, C. P., Alexandroff, R., Allende Prieto, C., et al. 2012, \apjs, 203, 21

\bibitem[Albert et al.(2011)]{alb11}
Albert, L., \'{E}tienne, A., Delorme, P., et al. 2011, \aj, 141, 203

\bibitem[Appenzeller et al.(1998)]{app98}
Appenzeller, I., Fricke, K., Furtig, W., et al. 1998, The Messenger, 94, 1

\bibitem[Beichman et al.(2013)]{bei13}
Beichman, C., Gelino, C. R., Kirkpatrick, J. D., et al. 2013, \apj, 764, 101

\bibitem[Beichman et al.(2014)]{bei14}
Beichman, C., Gelino, C. R., Kirkpatrick, J. D., et al. 2014, \apj, 783, 68


\bibitem[Burgasser et al.(2004)]{bur04}
Burgasser, A. J., McElwain, M. W., Kirkpatrick, J. D., et al. 2004, \aj, 127,
2856



\bibitem[Burningham et al.(2010)]{burn10}
Burningham, B., Pinfield, D. J., Lucas, P. W., et al. 2010, \mnras, 406, 1885


\bibitem[Burrows et al.(2003)]{bur03}
Burrows, A., Sudarsky, D., \& Lunine, J. I. 2003, \apj, 596, 587

\bibitem[Cushing et al.(2011)]{cus11}
Cushing, M. C., Kirkpatrick, J. D., Gelino, C. R., et al. 2011, \apj, 743, 50

\bibitem[Cushing et al.(2014)]{cus14}
Cushing, M. C., Kirkpatrick, J. D., Gelino, C. R., et al. 2014, \aj, 147, 113

\bibitem[Delorme et al.(2008)]{del08}
Delorme, P., Delfosse, X., Albert, L., et al. 2008, \aap, 482, 961


\bibitem[Dupuy \& Kraus(2013)]{dup13}
Dupuy, T. J., \& Kraus, A. L. 2013, Science, 341, 1492

\bibitem[Dupuy \& Liu(2012)]{dup12}
Dupuy, T. J., \& Liu, M. C. 2012, \apjs, 201, 19


\bibitem[Kimble et al.(2008)]{kim08}
Kimble, R. A., MacKenty, J. W., O'Connell, R. W., \& Townsend, J. A. 2008,
Proc. SPIE, 7010, 43


\bibitem[Kirkpatrick et al.(2013)]{kir13}
Kirkpatrick, J. D., Cushing, M. C., Gelino, C. R., et al. 2013, \apj, 776, 128

\bibitem[Kirkpatrick et al.(2012)]{kir12}
Kirkpatrick, J. D., Gelino, C. R., Cushing, M. C., et al. 2012, \apj, 753, 156

\bibitem[Kissler-Patig et al.(2008)]{kis08}
Kissler-Patig, M., Pirard, J.-F., Casali, M., et al. 2008, \aap, 491, 941

\bibitem[Krist(1995)]{kri95}
Krist, J. 1995, ASP Conf. Ser. 77, Astronomical Data Analysis Software and
Systems IV, ed. R.A. Shaw, H.E. Payne, \& J.J.E. Hayes (San Francisco,
CA: ASP), 349 



\bibitem[Leggett et al.(2010)]{leg10a}
Leggett, S. K., Burningham, B., Saumon, D., et al. 2010, \apj, 710, 1627

\bibitem[Leggett et al.(2013)]{leg13}
Leggett, S. K., Morley, C. V., Marley, M. S., et al. 2013, \apj, 763, 130


\bibitem[Liu et al.(2011)]{liu11}
Liu, M. C., Deacon, N. R., Magnier, E. A., et al. 2011, \apj, 740, 108

\bibitem[Liu et al.(2012)]{liu12}
Liu, M. C., Dupuy, T. J., Bowler, B. P., Leggett, S. K., \& Best, W. M. J.
2012, \apj, 758, 57

\bibitem[Lodieu et al.(2013)]{lod13}
Lodieu, N., B\'{e}jar, V. J. S., \& Rebolo, R. 2013, \aap, 550, L2

\bibitem[Lucas et al.(2010)]{luc10}
Lucas, P. W., Tinney, C. G., Burningham, B., et al. 2010, \mnras, 408, L56

\bibitem[Luhman(2014)]{luh14}
Luhman, K. L. 2014, \apj, 786, L18

\bibitem[Luhman et al.(2011)]{luh11}
Luhman, K. L., Burgasser, A. J., \& Bochanski, J. J. 2011, \apj, 730, L9

\bibitem[Luhman et al.(2012)]{luh12}
Luhman, K. L., Burgasser, A. J., Labb\'e, I., et al. 2012, \apj, 744, 135



\bibitem[Marsh et al.(2013)]{mar13}
Marsh, K. A., Wright, E. L., Kirkpatrick, J. D. et al. 2013, \apj, 762, 119

\bibitem[Morley et al.(2012)]{mor12}
Morley, C. V., Fortney, J. J., Marley, M. S., et al. 2012, \apj, 756, 172

\bibitem[Morley et al.(2014)]{mor14}
Morley, C. V., Marley, M. S., Fortney, J. J., et al. 2014, \apj, 787, 78

\bibitem[Pinfield et al.(2014)]{pin14}
Pinfield, D. J., Gromadzki, M., Leggett, S. K., et al. 2014, \mnras, in press

\bibitem[Saumon \& Marley(2008)]{sau08}
Saumon, D., \& Marley, M. S. 2008, \apj, 689, 1327

\bibitem[Saumon et al.(2012)]{sau12}
Saumon, D., Marley, M. S., Abel, M., Frommhold, L., \& Freedman, R. S.
2012, \apj, 750, 74



\bibitem[Simons \& Tokunaga(2002)]{sim02}
Simons, D. A., \& Tokunaga, A. 2002, \pasp, 114, 169

\bibitem[Subasavage et al.(2009)]{sub09}
Subasavage, J. P., Jao, W.-C., Henry, T. J., et al. 2009, \aj, 137, 4547

\bibitem[Tinney et al.(2012)]{tin12}
Tinney, C. G., Faherty, J. K., Kirkpatrick, J. D., et al. 2012, \apj, 759, 60

\bibitem[Tokunaga et al.(2002)]{tok02}
Tokunaga, A. T., Simons, D. A., \& Vacca, W. D. 2002, \pasp, 114, 180

\bibitem[Tokunaga \& Vacca(2005)]{tok05}
Tokunaga, A. T., \& Vacca, W. D. 2005, \pasp, 117, 421

\bibitem[Warren et al.(2007)]{war07}
Warren, S. J., Mortlock, D. J., Leggett, S. K., et al. 2007, \mnras, 381, 1400

\bibitem[Werner et al.(2004)]{wer04}
Werner, M. W., Roellig, T. L., Low, F. J., et al. 2004, \apjs, 154, 1

\bibitem[Wright et al.(2010)]{wri10}
Wright, E. L., Eisenhardt, P. R. M., Mainzer, A. K., et al. 2010, \aj, 140, 1868

\end{thebibliography}
\end{document}